\input{aipcheck}

\documentclass[
    ,final            
  ]
  {aipproc}

\layoutstyle{6x9}

\usepackage{amsmath}
\begin{document}

\title{The large$-N$ behavior of the holographic
models of QCD\footnote{Talk given at the $2^{nd}$ annual
\emph{France China Particle Physics Laboratory} (FCPPL) Workshop,
March 21-24, 2009, Wuhan, China.}}


\author{Fr\'ed\'eric Jugeau}{
  address={Institute of High Energy Physics, Chinese Academy of Sciences,\\
   P. O. Box 918(4), Beijing 100049, China.\\
Theoretical Physics Center for Science Facilities, Chinese Academy of Sciences,\\
 Beijing 100049, China (jugeau@ihep.ac.cn).}}

\begin{abstract}
The $AdS$/QCD dictionary is considered, checking the large$-N$
behavior of $5d$ dual models of QCD as a guideline. Especially, a
consistent chiral symmetry breaking function in the Hard Wall
model is derived and the different forms of the field/operator
correspondence are emphasized.
\end{abstract}

\maketitle

\section{Introduction}
There is a new hope that the $AdS$/CFT correspondence provides us
a new way to deal with the non-perturbative regime of QCD. Any
attempt to establish the $AdS$/QCD duality starting from the
tested $AdS$/CFT recipe \cite{Maldacena AdS/CFT} should then lead
to a consistent large$-N$ behavior for the $5d$ holographic models
of QCD, the tree level approximation of the dual string theory
corresponding indeed to the 't Hooft limit of the gauge theory.
Within the \emph{bottom-up} approach, the so-called Hard Wall
model \cite{PT} consists in breaking the underlying conformal
symmetry by truncating the holographic $AdS_5$ space-time such as
$0<z\leq z_m\simeq1/\Lambda_{QCD}$. Moreover, one usually assumes
that only a small number of QCD operators are relevant for
describing the low-energy phenomenology of the lightest mesons.
Among them, the scalar bulk field $v(z)$ dual to the quark
condensate $\langle\overline{q}q\rangle$ accounts for the chiral
symmetry breaking (in the sequel, we shall work with units in
which the $AdS_5$ radius is put equal to 1):
\begin{equation}\label{UV boundary condition of v}
v(z)=z(m_q+\overline{\sigma}z^2)\underset{z\to0}{\to}m_q z\;\;.
\end{equation}
The source field $m_q$ is nothing else than the light quark mass
while $\overline{\sigma}$ is rather systematically identified in
the literature with the chiral condensate
$-\langle\overline{q}q\rangle$. As pointed out in \cite{CCW}, such
a naive identification leads to inconsistencies regarding the
large$-N$ behavior since $m_q\sim O(N^0)$ and
$\langle\overline{q}q\rangle\sim O(N)$. The aim of this talk is
thus to reexamine the $AdS$/QCD dictionary, checking the large$-N$
behavior of the Hard Wall model as a guideline. Doing so, we shall
be able to identify unambiguously the quark mass and the chiral
condensate entering the expression of the so-called chiral
symmetry breaking function $v(z)$ \cite{jugeau}.

\section{The chiral symmetry breaking function and the $AdS$/QCD
dictionary}

The $5d$ holographic space-time is described by the following
conformally flat metric ($M,N=0,\ldots,4$ and
$\mu,\nu=0,\ldots,3$):
\begin{equation}\label{IR hard wall metric}
ds^2=g_{MN}(z)dx^Mdx^N=e^{2A(z)}\Big(\eta_{\mu\nu}dx^{\mu}dx^{\nu}-dz^2\Big)
\end{equation}
where $\eta_{\mu\nu}=\textrm{diag}(+1,-1,-1,-1)$ is the flat
metric tensor of the $4d$ Minkowski boundary space, $A(z)=-\ln(z)$
is the warp factor and $0<z\leq z_m$. The effective action able to
reproduce the chiral properties of QCD is
\begin{equation}\label{action}
S_{5d}=-\frac{1}{2\kappa^2}\int
d^5x\sqrt{g}\,\Big(\mathcal{R}+\Lambda\Big)+\frac{1}{k}\int d^5
x\sqrt{g}\,Tr\,\Big\{|DX|^2-m_5^2\,X^2-\frac{1}{4g_5^2}(G_L^2+G_R^2)\Big\}
\end{equation}
with $g=\textrm{det}(g_{MN})$ the determinant of the metric
tensor, $\kappa^2\equiv8\pi G_{5}$ the $5d$ Newton constant and
where the $AdS_5$ Ricci scalar
$\mathcal{R}=-\frac{5}{3}\Lambda=20$ is, according to our
convention \eqref{IR hard wall metric}, positive \emph{i.e.} the
cosmological constant $\Lambda$ is negative. The overall parameter
$k$ has the dimension of a length while the $5d$ gauge coupling
constant $g^2_5$ is dimensionless. The relevant QCD operators are
the $SU(2)_L\times SU(2)_R$ left-handed and right-handed currents
$j_{L\,\mu}^a=\overline{q}_L\gamma_{\mu}t^a q_L$ and
$j_{R\,\mu}^a=\overline{q}_R\gamma_{\mu}t^a q_R$ and the quark
operator $\overline{q}_R^iq_L^j$ (when $n_F=2$,
$t^a=\frac{\sigma^a}{2}$ are the Pauli matrices with $a=1,2,3$ and
$i,j=1,2$). The bulk field content of the $5d$ dual theory
consists then of the left-handed and right-handed gauge fields
$A_{L_M}^a(x,z)$ and $A_{R_M}^a(x,z)$ and of the bi-fundamental
scalar field:
\begin{equation}\label{scalar bulk field}
X^{ij}(x,z)=\Big(\frac{v(z)}{2}+S(x,z)\Big)^{ik}\Big(e^{2i\pi^b(x,z)t^b}\Big)^{kj}
\end{equation}
which is massive $m_5^2=(\Delta-p)(\Delta+p-4)=-3$ with $\Delta=3$
and $p=0$ and involves the bulk field $S(x,z)$ dual to the scalar
mesons, the dimensionless pseudo-scalar field $\pi(x,z)$ and the
background field $v(z)$, written as $v^{ik}(z)=v(z)\delta^{ik}$ in
the isospin limit. The linearized equations of motion for the
different bulk fields can be derived straightforwardly from the
quadratic part of the $5d$ effective action \eqref{action}. In
particular, the ($4d$ Fourier transforms of the) pseudo-scalar
modes are not independent but satisfy two coupled differential
equations:
\begin{equation}\label{HW pion EOM 2}
\partial_z\Big(\frac{1}{z}\partial_z\tilde{\phi}^a(q,z)\Big)-\frac{g_5^2v(z)^2}{z^3}\Big(\tilde{\phi}^a(q,z)-\tilde{\pi}^a(q,z)\Big)=0\;\;,
\end{equation}
\begin{equation}\label{HW pion EOM 1}
q^2\partial_z\tilde{\phi}^a(q,z)-\frac{g_5^2v(z)^2}{z^2}\partial_z\tilde{\pi}^a(q,z)=0
\end{equation}
where $\phi$ contributes to the longitudinal part of the
axial-vector bulk field $A_M^a=(A^a_L-A^a_R)/2$, defined as
$\partial\phi=A-A_{\perp}$.

According to the Gell-Mann-Oakes-Renner (GMOR) relation
$m_{\pi}^2f_{\pi}^2=2m_q\sigma+O(m_{q}^2)$, the magnitudes of the
pion and quark masses are related to the size of the chiral
condensate $\sigma\equiv-\langle\overline{q}q\rangle$. In the
chiral limit, the massless pions are the pseudo-Goldstone bosons
associated with the spontaneous breaking of the chiral symmetry.
They acquire a mass provided that the chiral symmetry is
explicitly broken. On the other hand, the chiral symmetry breaking
mechanism is holographically described by the background field
$v(z)$ which reads in the Hard Wall model
$v(z)=z(\overline{m}+\overline{\sigma}z^2)$. The two parameters
$\overline{m}$ and $\overline{\sigma}$, which do not have to be
identified \emph{a priori} with the quark mass and the chiral
condensate accounts for the explicit and implicit chiral symmetry
breaking respectively and, as such, should be naturally
proportional, on the one hand, to $m_q$ and, on the other hand, to
$\sigma$. It is worth deriving shortly the GMOR relation. When
$q^2=m_{\pi}^2$, Eq.\eqref{HW pion EOM 1} reads
\begin{equation}\label{normalizable mode relation}
m_{\pi}^2\partial_z\phi(z)=\frac{g^2_5v(z)^2}{z^2}\partial_z\pi(z)\;\;.
\end{equation}
The normalizable modes of the pions behave as $\phi(0)=0$,
$\partial_z\phi(z)\big|_{z_m}=0$ and $\pi(0)=0$ while its IR
boundary condition is dictated by the chiral symmetry (the
subscript $\chi$ refers to the chiral limit): since
$\pi_{\chi}(z)=-1$ is solution of Eq.\eqref{HW pion EOM 1} when
$q^2=0$, $\pi(z)$ must approach $-1$ at low energies. Then, at the
leading order, Eq.\eqref{normalizable mode relation} becomes:
\begin{equation}\label{GMOR2}
\pi(z)=m_{\pi}^2\int_0^z du f(u)\frac{1}{g_5^2
u}\partial_u\phi_{\chi}(u)
\end{equation}
where the function $f(u)\equiv\frac{u^3}{v(u)^2}$ exhibits a peak
located at $u_c=\sqrt{\frac{\overline{m}}{3\overline{\sigma}}}$.
So, only small values of $u$ give sizeable contributions to the
integral provided that $\overline{m}$ is small in comparison with
$\overline{\sigma}$, which is consistent with the fact that
$\overline{m}$ and $\overline{\sigma}$ are related to $m_q$ and
$\sigma$ respectively. The remaining function can then be
evaluated at $u\simeq0$ which yields the pion decay constant
$f_{\pi}^2=-\frac{1}{kg_5^2}\,\frac{1}{z}\partial_z\phi_{\chi}(z)\big|_{z=0}$.
Finally, in order to recover the GMOR relation at leading order,
one takes the large $z$ limit which gives rise to the simple
condition $\overline{m}\,\overline{\sigma}=k m_q\sigma$ (9). It is
then consistent to take $\overline{m}=m_q$ (10) and
$\overline{\sigma}=k \sigma$ (11). On the other hand, since any
holographic model must predict the same high energy physics near
the UV brane (when $z\to0$), the same value of $k$ should be
extracted within any dual model of QCD. The short distance
$AdS$/QCD matching of the two-point correlation function of the
scalar current gave, in the framework of the Soft Wall model, the
value $k=\frac{16\pi^2}{N}$ \cite{soft wall scalar}. The chiral
symmetry breaking function takes then the following form:
\begin{equation}\label{v1}
v(z)=z\Big(m_q+\frac{16\pi^2}{N}\sigma
z^2\Big)\underset{z\to0}{\to}m_q z\;\;.\tag{12}
\end{equation}
Its large$-N$ scaling is consistent since $m_q\sim O(N^0)$ and
$\sigma\sim O(N)$ yield $v(z)\sim O(N^0)$.

Let us now clarify the $AdS$/QCD dictionary. A mere rescaling of
the scalar bulk field $X(x,z)$ and of the $5d$ coupling constant :
\begin{equation}\label{redefinition}
\hat{X}(x,z)=\frac{a}{2}\,X(x,z)\;\;\;\textrm{and}\;\;\;\frac{1}{\hat{g}_5^2}=\frac{a^2}{4}\,\frac{1}{g_5^2}\;\;,\tag{13}
\end{equation}
where the parameter $a=\sqrt{N}/2\pi$ has been introduced in
\cite{CCW}, implies a renormalization of the $5d$ holographic
action \eqref{action}:
\begin{equation}\label{renormalization}
\frac{1}{k}\int
d^5x\sqrt{g}\,Tr\,\mathcal{L}(X,g_5^2,A_L,A_R)\;\;\;\Rightarrow\;\;\;\int
d^5x\sqrt{g}\,Tr\,\mathcal{\hat{L}}(\hat{X},\hat{g}_5^2,A_L,A_R)\;\;.\tag{14}
\end{equation}
The action, written in the first form, involves an overall factor
$1/k$ related to the number of colors $N$. Because of the scalar
bulk field and gauge coupling redefinition, this $N$-dependent
factor is absorbed on the \emph{r.h.s.} of (14). That implies
modifications in the field/operator prescription, the chiral
symmetry breaking function becoming indeed:
\begin{equation}\label{v2}
\hat{v}(z)=\frac{a}{2}\,m_q z+\frac{2}{a}\,\sigma
z^3\underset{z\to0}{\to}\frac{a}{2} m_q z\;\;.\tag{15}
\end{equation}
We have then $\overline{m}=\frac{a}{2}\,m_q$ and
$\overline{\sigma}=\frac{2}{a}\,\sigma$ which fulfils the
condition $\overline{m}\,\overline{\sigma}=m_q\sigma$, analogous
to (9). Moreover, the large$-N$ behavior has changed, being now
$\hat{v}(z)\sim O(\sqrt{N})$. Besides, one usually chooses to
renormalize the $5d$ gauge coupling such that
$\hat{g}_5^2=12\pi^2/N$ instead of the gauge bulk fields
$A_{L_M}^a$ and $A_{R_M}^a$ (or, equivalently, $V_M^a$ and
$A_M^a$) which has the advantage of letting unchanged the boundary
conditions:
\begin{equation}
\tilde{V}(q^2,0)=\tilde{A}(q^2,0)=1\;\;\;\;\;\;\;\;\textrm{and}\;\;\;\;\;\;\;\;
\partial_z\tilde{V}(q^2,z)\big|_{z=z_m}=\partial_z\tilde{A}(q^2,z)\big|_{z=z_m}=0.\tag{16}
\end{equation}
Also, the pseudo-scalar modes are not affected by the rescaling
(13). As a result, all the $5d$ equations of motion and the
boundary conditions, especially $\pi_{\chi}(z)=-1$, remain the
same with $g_5^2$ and $v(z)$ replaced by $\hat{g}_5^2$ and
$\hat{v}(z)$ such that the products $g_5^2\,v(z)^2$ and
$\hat{g}_5^2\,\hat{v}(z)^2$, as in \eqref{HW pion EOM 2}-\eqref{HW
pion EOM 1} for instance, are both $N$-independent in the chiral
limit. As for the vector, axial-vector and pseudo-scalar decay
constants, which read respectively
\begin{equation}
F_{V_n}^2=\frac{1}{kg_5^2}\Big(\frac{1}{z}\partial_z
v_n(z)\Big)^2\Big|_{z=0}\;\;\;\;,\;\;\;\;
F_{A_n}^2=\frac{1}{kg_5^2}\Big(\frac{1}{z}\partial_z
a_n(z)\Big)^2\Big|_{z=0}\;\;\;\;,\;\;\;\;
f_{\pi}^2=-\frac{1}{kg_5^2}\,\frac{1}{z}\partial_z\phi_{\chi}(z)\Big|_{z=0}\tag{17}
\end{equation}
with
$v_n(z)=\sqrt2\frac{z}{z_m}\frac{J_1(m_{V_n}z)}{J_1(m_{V_n}z_m)}$
and $a_n(z)$ the vector and axial-vector normalizable wave
functions subjects to vanishing UV and IR boundary conditions,
they retain their $AdS$ expressions except for the overall factor
$1/k g_5^2$ which becomes $1/\hat{g}_5^2$.

\section{The large$-N$ behavior of the Hard Wall model}

After having determined unambiguously the light quark mass and the
chiral condensate entering the expression of the chiral symmetry
breaking function (12) and clarified the $AdS$/QCD dictionary as
used in the \emph{bottom-up} approach of the holographic models of
QCD, let us focus on the large$-N$ behavior of the Hard Wall
model. A glance at the bulk fields, be it, for example, the
normalizable vector mode $v_n(z)$ above or the vector
bulk-to-boundary propagator for space-like momentum
($Q\equiv\sqrt{-q^2}>0$):
\begin{equation}
\tilde{V}(Q^2,z)=Qz\Big(K_1(Qz)+\frac{K_0(Qz_m)}{I_0(Qz_m)}I_1(Qz)\Big)\;\;,\tag{18}
\end{equation}
leads us to conclude that they do not scale with $N$. As a
consequence, the square of the decay constants $F_{V_n}^2$,
$F_{A_n}^2$ and $f_{\pi}^2$ in (17) scale as $O(N)$ which is
reminiscent of what happens in Chiral Perturbation Theory with the
large$-N$ behavior implemented \cite{Pich}. Let us go further by
checking the large$-N$ behavior of the electromagnetic
$F_{\pi}(q^2)$ and gravitational $A_{\pi}(q^2)$ form factors of
the pion \cite{rad} which take, in the chiral limit, very similar
forms:
\begin{equation}
F_{\pi}(Q^2)=\frac{1}{kg_5^2}\frac{1}{f_{\pi}^2}\int_0^{z_m}dz\,z\,
\tilde{V}(Q^2,z)\Big[\Big(\frac{\partial_z\Psi(z)}{z}\Big)^2+\frac{g_5^2v(z)^2}{z^4}\Psi(z)^2\Big]\;\;,\tag{19}
\end{equation}
\begin{equation}
A_{\pi}(Q^2)=\frac{1}{kg_5^2}\frac{1}{f_{\pi}^2}\int_0^{z_m}dz\,z
\,\tilde{h}(Q^2,z)\Big[\Big(\frac{\partial_z\Psi(z)}{z}\Big)^2+\frac{g_5^2v(z)^2}{z^4}\Psi(z)^2\Big]\;\;.\tag{20}
\end{equation}
The so-called chiral field
$\Psi(z)\equiv\phi_{\chi}(z)-\pi_{\chi}(z)$ is associated with the
massless pseudo-scalar mode while $\tilde{h}(Q^2,z)$ is the tensor
bulk-to-boundary propagator, analogous to (18), related to the
fluctuations around the $AdS_5$ conformally flat metric \eqref{IR
hard wall metric}. Besides their charge normalizations
$F_{\pi}(0)=A_{\pi}(0)=1$, the form factors do not scale with $N$
which stems from the compensation mechanism between $1/kg_5^2\sim
O(N)$ and $1/f_{\pi}^2\sim O(1/N)$. As for the $VPP$ coupling
constants, they follow writing the vector bulk-to-boundary
propagator in terms of the vector mass poles $m_{V_n}$, of the
residues $F_{V_n}$ and of the normalizable eigenfunctions
$v_n(z)$:
\begin{equation}
\tilde{V}(q^2,z)=-\sqrt{kg_5^2}\sum_{n=1}^{\infty}\frac{F_{V_n}\,v_n(z)}{q^2-m_{V_n}^2+i\epsilon}\;\;.\tag{21}
\end{equation}
which does not scale, as it must be, with $N$. The $AdS$
expression of the vector form factor and of the the
$g_{V_n\pi\pi}$ couplings follow:
\begin{equation}
F_{\pi}(q^2)=-\sum_{n=1}^{\infty}\frac{F_{V_n}\,g_{V_n\pi\pi}}{q^2-m_n^2+i\epsilon}\tag{22}
\end{equation}
with
\begin{equation}
g_{V_n\pi\pi}=\frac{1}{\sqrt{kg_5^2}}\frac{1}{f_{\pi}^2}\int_0^{z_m}dz\,z\,v_n(z)\,\Big[\Big(\frac{\partial_z\Psi(z)}{z}\Big)^2+\frac{g_5^2v(z)^2}{z^4}\Psi(z)^2\Big]\;\;.\tag{23}
\end{equation}
As a result, the large$-N$ counting rules are consistent in the
$AdS$/QCD Hard Wall model as $g_{V_n\pi\pi}\sim O(1/\sqrt{N})$
such that we recover free, stable, non-interacting mesons at large
$N$. Moreover, imposing that the pion decay constant coincides
with the value $f_{\pi}\simeq92.4$ MeV in the chiral limit, one
gets, with $1/z_m\simeq323$ MeV, the quark condensate
$\sigma\simeq(171\;\textrm{MeV})^3$. Because of the factor of $k$
in the definition of $\overline{\sigma}=k\sigma$, the condensate
turns out to be noticeably smaller than Hard Wall model estimates
where this factor is usually omitted.

\section{Conclusion}

This talk aimed at clarifying some confusion regarding the
large$-N$ 't Hooft limit of the Hard Wall model of QCD which, used
as a guideline, led us to study the different forms of the
field/operator prescription within $AdS$/QCD. Such subtleties are
not systematically taken into account when dealing with different
$5d$ holographic models.



\bibliographystyle{aipproc}   


\IfFileExists{\jobname.bbl}{}
 {\typeout{}
  \typeout{******************************************}
  \typeout{** Please run "bibtex \jobname" to optain}
  \typeout{** the bibliography and then re-run LaTeX}
  \typeout{** twice to fix the references!}
  \typeout{******************************************}
  \typeout{}
 }


\end{document}